# Realized Stable BP-N at Ambient Pressure by Phosphorus Doping

Guo Chen,[1,2] Chengfeng Zhang[1,2], Yuanqin Zhu[1,2], Bingqing Cao[1,2], Jie Zhang[1], and Xianlong Wang[1,2*]

[1]*Key Laboratory of Materials Physics, Institute of Solid State Physics, HFIPS, Chinese Academy of Sciences, Hefei 230031, China*

[2]*University of Science and Technology of China, Hefei 230026, China*

───────────

[*]Author to whom all correspondence should be addressed: xlwang@theory.issp.ac.cn

# ABSTRACT

Black phosphorus nitrogen (BP-N) is an attractive high-energy-density material. However, high-pressure synthesized BP-N will decompose at low-pressure and cannot be quenched to ambient conditions. Finding a method to stabilize it at 0 GPa is of great significance for its practical applications. However, unlike cg-N, LP-N, and HLP-N, it is always a metastable phase at high-pressure up to 260 GPa, and decomposes into chains at 23 GPa. Here, based on the first-principles simulations, we find that P atom doping can effectively reduce the synthesis pressure of BP-N and maintain its stability at 0 GPa. Uniform distribution of P atom dopants within the layer helps maintain the structural stability of BP-N layer at 0 GPa, while interlayer electrostatic interaction induced by N-P dipoles enhances its dynamic stability by eliminating interlayer slipping. Furthermore, pressure is conducive to enhancing the stability of BP-N and its doped forms by suppressing N-chain dissociation. For the configuration with 12.5% doping concentration, a gravimetric energy density of 8.07 kJ/g can be realized, which is nearly two times higher than TNT.

**INTRODUCTION**

Metastable allotropes of molecular and nonmolecular nitrogen have been extensively investigated as potential high-energy-density materials (HEDMs) for future applications. This interest stems from the fact that the N≡N triple bond in $N_2$ is one of the strongest chemical bonds found in molecules, with a bonding energy of 954 kJ/mol. Transforming this triple bond into a double bond (N=N, bonding energy of 418 kJ/mol) or a single bond (N−N, bonding energy of 160 kJ/mol) significantly decreases the bonding energy and increases the total energy [1]. Conversely, decomposing N−N or N=N into N≡N results in the release of a tremendous amount of energy. Consequently, polymerized nitrogen containing N−N or N=N bonds has attracted significant interest as HEDMs.

Nonetheless, due to the strength of the triple bond and the fragility of the single bond, synthesizing polymerized nitrogen at ambient pressure is challenging [2]. Direct compression of molecular nitrogen or nitrogen-containing compounds (e.g., azides) often leads to pressure-induced symmetry breaking phases [3–11], thus expanding nitrogen's structural diversity. Theoretically, numerous polymerized nitrogen structures, such as rings [12], layers [13–16], clusters, chains [17–19], cages [20], ions [21], and polynitrogen compounds [22–32], have been predicted. Scientists have been attempting to synthesize polymerized nitrogen and polynitrogen compounds since the first isolation of nitrogen gas in 1772 and the first synthesis of azide anions in 1890. For instance, on chemical methods, the $N_5^+$ ion was first synthesized in 1999 [2]. And the synthesis of $N_5^-$ has several breakthrough processes since 2016 [33]. Alternative, high-pressure have proven as an effective and popular methods for synthesizing polymerized nitrogen. The first polymeric all-nitrogen materials, cg-N, was synthesized at 110 GPa in 2004 [4]. LP-N and HLP-N were subsequently synthesized at the pressures of 120 GPa and 180 GPa [34,35], respectively. Remarkably, prior to 2020, nitrogen had never been observed in an isostructural crystal structure similar to black phosphorus (BP), which possesses unique puckered two-dimensional layers [36]. However, in 2020, a significant breakthrough was achieved by successfully synthesizing BP-N at a pressure of 146 GPa for the first time [37–39]. This groundbreaking achievement completed the set of BP structures for Group V elements.

However, high-pressure synthesized BP-N decomposed with pressure decreasing to 48 GPa [37]. Similar decomposition phenomena at low-temperature have been observed in other high-pressure synthesized polymeric all-nitrogen forms, such as cg-N [4], LP-N [34], and HLP-N [35], which decompose at 42 GPa, 52 GPa, and 66 GPa, respectively. The instability of polymeric all-nitrogen materials at low-pressure greatly limits its potential applications as HEDMs. Strategies to enhance the stability of polymerized nitrogen include forming compounds with alkali or alkaline earth metals to not only discover new polymerized nitrogen types but also decrease its synthesis pressure [40–49]. Doping is another effective method [50,51], as it can redistribute the electrons in the system, and the dopants can act as bonding agents. Moreover, employing nano-confinement effects and surface modification can further improve its stability [52–57]. Our previous research demonstrated that cg-N's decomposition at low-pressure was caused by surface instability, and surface modification can realize stable cg-N at 0 GPa [58].

Unlike other experimental synthesized polymeric all-nitrogen materials, BP-N demonstrates an unstable configuration that decomposes into chains at low-pressure's structural relaxation calculation [37]. To obtain BP-N at 0 GPa, first, we should maintain the BP-N structure and prevent the formation of a chain-like structure at low-pressure, realizing structural stability. Subsequently, it is crucial to ensure the absence of imaginary frequencies in the phonon dispersions, realizing dynamical stability. Phosphorus and nitrogen are both in Group V of the periodic table and have similar electron structures and chemical properties. Interestingly, BP demonstrates remarkable stability at ambient conditions [59]. Moreover, BP exhibits promising properties and holds significant potential for various applications [36]. As a result, phosphorus doping introduces a promising avenue for achieving stability of BP-N at low-pressure, which shares phase similarities with BP.

In this work, we investigated the stability of P atom doped BP-N at high-pressure with different doping concentrations, doping uniformity, and doping configurations using the first-principles method. Our results shown that P atom doping plays a crucial role in maintaining the stability of BP-N. Uniform distribution of dopants in the BP-N layer helps maintain its structural stability at 0 GPa, while interlayer electrostatic interactions due to N-P dipoles enhance dynamic stability by preventing interlayer slipping.

**METHODS**

The first-principles calculations of the structure and electronic properties based on the density functional theory with a plane-wave basis set were performed using the Vienna *ab initio* simulation package (VASP) [60]. The generalized gradient approximation (GGA) [61,62] parameterized by Perdew-Burke-Ernzerhof (PBE) [63] exchange-correlation functional, and projected augmented wave (PAW) [64] with N-$2s^22p^3$ and P-$3s^23p^3$ as valence electrons were used for calculation. The atomic positions, lattice parameters, and cell volume were fully optimized using the conjugate gradient (CG) algorithm and the iteration relaxation of the atomic positions stopped when all forces were less than 0.001 eV/Å. The convergence of total energy and stress tensor was set to $1\times10^{-6}$ eV and 0.01 GPa, respectively. The Monkhorst-Pack method was used to analyze the reciprocal space of all Brillouin zones. The k-grid spacing was set to 0.20 Å$^{-1}$, ensuring a K-point density that converged the energy within 0.5 meV/atom. The phonon dispersions were calculated using the phonopy code based on the finite displacement method. The PBE-vdW method was used for all structures, which included van der Waals corrections for the interlayer interaction [65]. To compare the enthalpies of investigated polymerized nitrogen, supercells containing 32 atoms were used to ensure the comparability among BP-N, cg-N, LP-N, and HLP-N. Other calculations about P atom doped BP-N are performed using a $1\times1\times2$ supercell with 16 atoms. For 12.5% doping concentration, all possible configurations were investigated, while for the 25% doping concentration, we randomly selected 50 non-identical configurations. For the calculation of the decomposition enthalpy and energy density of BP-N, α'-$P_3N_5$ [66] and α-$N_2$ [27] were chosen as the ground state phase. VAPKIT [67] and MULTIWFN [68] are also adopted in our work.

The uniformity of P atoms distribution was calculated using the following equation:

$$\text{Uniformity} = 1 - \frac{\sum_{i=1}^{n}\sum_{j=1,i\neq j}^{n}\frac{1}{\sqrt{2\pi}}e^{-\frac{(x_i-x_j)^2}{2}}}{n} \tag{1}$$

Where x, n, i, and j are the cartesian position of P atoms (Å), the total number of P atoms, and the indices for the i-th and j-th P atoms. The distances between these P atoms are used as criteria for evaluating uniformity based on gauss distribution. The value of the gaussian distribution increases when the no.i and no.j P atoms are closer, indicating a lower level of uniformity.

The Kamlet–Jacobs empirical equations for detonation speed ($V_d$) and detonation pressure ($P_d$) are as follows [69]:

$$V_d = 1.01(NM^{0.5}E_d^{0.5})^{0.5}(1 + 1.30\rho) \tag{2}$$

$$P_d = 15.58\rho^2 NM^{0.5}E_d^{0.5} \tag{3}$$

Where $N$, $M$, $E_d$, and $\rho$ represent moles of $N_2$ per gram of explosives (mol/g), the molar mass for $N_2$ gas (28 g/mol), energy density (J/g), and the density (g/cm³), respectively.

**RESULTS AND DISCUSSION**

To investigate the decomposition mechanism of BP-N at low-pressure, we fully relaxed the BP-N structure with pressure gradually decreasing from 100 GPa. In Fig. 1(a), the variation of longest and average bond lengths of BP-N is shown as a function of pressure. Greater than 25 GPa, there is a gradual and smooth change in bond length variations, and the BP-N structure is still maintained. Moreover, it has an average bond length close to 1.50 Å at 25 GPa, which is larger than the bond length of cg-N at 0 GPa (1.40 Å) [70,71]. However, when the pressure falls below 25 GPa, the bond lengths between nitrogen increase drastically, indicating a phase transition occurs near this pressure. As shown in Fig. 1(b) and Fig. 1(c), the BP-N will decompose to nitrogen zigzag chains at the pressure of 23 GPa.

The phonon dispersions of BP-N structure from 100 GPa to 30 GPa were also calculated, and corresponding results at 40 GPa and 30 GPa are shown in Fig. 1(d) and Fig. 1(e) respectively. We found that beyond 40 GPa, BP-N illustrates the dynamic stability, however, its phonon dispersions at 30 GPa exhibits imaginary frequencies. The decomposition process of BP-N is summarized as following: With pressure decreasing, the BP-N will become dynamically unstable at 30~40 GPa, and then decomposes to nitrogen zigzag chains at 23 GPa [39].

Since BP exhibits exceptional stability at ambient conditions, we seek to use P atom doping to enhance the stability of the BP-N, which shares phase similarities with BP. The effects of P atom doping on the BP-N stability were investigated, and the enthalpy-pressure curve of different P atom doping concentrations for various phases, namely cg-N, BP-N, LP-N, and HLP-N as shown in Fig. 2. For pristine cases (Fig. 2(a)), the cg-N has the lowest enthalpy in the pressure up to the pressure of 186 GPa, where a phase transition from cg-N to LP-N occur. Finally, the HLP-N becomes the ground state at pressure higher than 280 GPa. The enthalpy of BP-N at high-pressure is always higher than that of cg-N, LP-N, and HLP-N, BP-N is the metastable phase in the pressure ranging from 60 GPa to 260 GPa, consistent with previous computational results [37–39].

As shown in Fig. 2(b), with 3.125% P atom doping concentration, the phase-transition sequence does not change, and only the phase-transition pressure is modified, for example, the phase-transition pressure of cg-N to LP-N decreases from 186 GPa to 140 GPa. Interestingly, after increasing the P atom doping concentration to 6.25%, as shown in Fig. 2(c), the phase-transition sequence is changed significantly, and a sequence of cg-N to BP-N to HLP-N is observed resulting in that the BP-N becomes the most stable phase at pressure ranging from 154 GPa to 234 GPa. The results indicate that P atom doping can enhance the thermodynamic stability of BP-N at high-pressure. Please note that, although BP-N exhibits a thermodynamically stable pressure region which spanning of 80 GPa at 6.125 % P atom doping concentration, it remains decomposed into N-chains at 0 GPa. Following, we discuss the effects of P atom doping on the BP-N stabilities with higher doping concentrations.

We constructed configurations with higher doping concentrations (12.5%, 25%). Fig. 3(a) illustrates the structural stability and uniformity of the various configurations at 0 and 50 GPa. We can find that in the cases of higher doping concentrations, more than half of P atom doping configurations can stabilize the BP-N structure at 0 GPa, indicating that P atom doping can prevent BP-N decomposition into N-chains. Furthermore, a distinct boundary differentiates stable and unstable structures, with uniformity serving as the key separator. Particularly noteworthy is the observation that structures exhibit instability when their uniformity falls below 0.80, especially in the case of 25% doping concentration. The instability observed can be attributed to the clustering of P atoms. Due to the significant difference in radius between P and N atom, large stress will be introduced between P-atom clustering regions and N-atom clustering regions, which is not conducive to structural stability. More importantly, without P atom doping to enhance stability, the regions with N-atom clustering will decompose. Moreover, the pressure is also conducive to stability, as evidenced by the greater number of configurations that stabilize at 50 GPa. In addition to that, we conducted an analysis on the relationship between the energy and uniformity of configurations with 12.5% and 25% P atom doping at 0 GPa. Our findings, depicted in Fig. 3(b), highlight that structures exhibiting higher uniformity tend to possess lower energy.

Since BP-N has a two-dimensional stacking structure, interlayer interactions may play a significant role in its dynamic stability, the investigation of the position for P atoms between the layers is necessary. A tendency towards uniform doping between layers are observed. In the case of 12.5% doping concentration, for all stable configurations, each layer contained an equal number of P atoms. Upon analyzing them, we found that their single-layer configurations were equivalent, with the only difference being the relative positions between the interlayer atoms. We calculated the energy of these configurations and studied their relationship with the distance of P atoms between the layers, as shown in Fig. 4(c). The configuration labeled by the red triangle which have the lowest energy, and it was observed the P atoms between the layers were arranged perpendicularly. Notably, it was the only one without imaginary frequencies. Its structure, along with phonon spectra, are shown in Fig. 4(a). Additionally, the configuration, indicated by a blue triangle in Fig. 4(c), which is one of those exhibiting imaginary frequencies in phonon dispersions, has its structure and phonon characteristics depicted in Fig. 4(b). For other structures with imaginary frequencies, they were similar to it, with the imaginary frequencies occurring at the Z point. The presence of imaginary frequencies at the Z point indicated that interlayer sliding has occurred, as revealed by the analysis of vibration modes, the black arrows on the figure indicate the direction of this vibration. In the configuration with the lowest energy, the lowest frequency at the Z point was observed at a certain distance away from zero. This signifies that this particular doping method is beneficial in preventing interlayer sliding. The strengthening of interlayer interactions can be attributed to the contribution of electrostatic interactions.

Due to the difference in electronegativity, there is a charge transfer from P atoms to N atoms, resulting in P atoms carrying a positive charge and N atoms carrying a negative charge. For example, in the configuration with the lowest energy (Fig. 4(a)), the average Bader charge of the P atom and N atom are 3.10 e/atom and 5.27 e/atom, respectively. In details, the N atom nearest to P atom has a Bader charge of 6.42 e, and as moving further away from the P atom, the charges of N atom decrease and approach 5.0 e. Furthermore, as shown in Fig. 4(d), for the N atoms which bonded to P atoms, they form the dipoles with the adjacent layer of P atoms, contributing to stable electrostatic interactions. However, in configurations with imaginary frequencies, the absence of these dipoles will cause interlayer sliding.

In the case of 25% doping concentration, we conducted an extensive analysis of all phonons for stable configurations and observed that there are four configurations without imaginary frequencies. In these configurations, the P atoms between the layers were arranged perpendicularly, and they all exhibit interlayer N-P dipoles. For the other configurations, imaginary frequencies appear at the Z point. The result is consistent with the case of 12.5% doping concentration. Fig. 5 depicts the two dynamically stable configurations, along with their respective phonon dispersions. In Fig. 5(a), a unit cell contains four dipoles and the lowest frequency at the Z point is 2.524 THz, while in Fig. 5(b), there are two dipoles and the lowest frequency at the Z point is 0.954 THz. When N-P dipoles decrease, the lowest frequency at the Z point will also decrease. It is worth mentioning that for configurations where P atoms between the layers are not arranged perpendicularly, we observed a tendency for these P atoms to align perpendicularly after structural relaxation in order to form more dipoles.

The decomposition enthalpies of 12.5% and 25% doped BP-N relative to α'-$P_3N_5$ [66] and α-$N_2$ [27] at ambient pressure are 1.34 and 1.18 eV/atom, respectively, indicating that they are potential high-energy-density materials. To further investigate their energy density and explosive performance, we conducted additional studies, experimental values for the known trinitrotoluene (TNT), 1,3,5,7-tetrazoctane (HMX) are as compared [72–75], as depicted in Fig. 6. Subsequent calculations revealed that gravimetric energy density ($E_d$) of 12.5% doping concentration (density: 3.06 g/cm$^3$) and 25 % doping concentration (density: 3.13 g/cm$^3$) are estimated to be approximately 8.07 and 6.27 kJ/g, respectively. These values are significantly higher than those of TNT and HMX. Notably, the $E_d$ of 12.5% doping concentration is approximately twice as high as that of TNT (4.3 kJ/g). The volumetric energy density ($E_v$) calculations further highlight the superior performance of 12.5% doping concentration, with values of 24.71 kJ/cm$^3$. This figure is more than three times higher than that of TNT (7.05 kJ/cm$^3$).

Furthermore, as illustrated in Fig. 6, we estimated the explosive performance of 12.5% and 25% doping concentrations by utilizing the Kamlet-Jacobs empirical equation [69]. This equation has been widely acknowledged for its validity. The detonation velocities ($V_d$) of 12.5 % and 25 % doping concentrations were calculated to be 18.07 and 15.02 km/s, respectively, while the corresponding pressures ($P_d$) were 1885.19 and 1314.39 kbar. Notably, 12.5% doping concentration exhibits $V_d$ values that two to three times higher than that of TNT (6.90 km/s) and $P_d$ values approximately ten times higher than that of TNT (190 kbar).

**CONCLUSION**

In conclusion, this research presents a systematic investigation into the stabilities of BP-N using first-principles calculations. It sheds light on the effects of various factors such as doping concentrations, doping uniformity, and doping configurations. P atom doping can induce BP-N to the ground state at high-pressure. Furthermore, a tendency towards uniform distribution of dopants can improve the structural stability of BP-N, which remains stable even at 0 GPa. By introducing P-atom dopants, the adhesion between the layers is strengthened through the interlayer electrostatic interactions of N-P dipoles, which reduces the tendency for interlayer slip and enhances the stability of the structure. Furthermore, Pressure is conducive to enhancing the stability of BP-N and its doped forms by preventing the decomposition into N-chain. The calculated $E_d$ ($E_v$) of 12.5% doped BP-N is 8.07 kJ/g (24.71 kJ/cm$^3$), which is about two (three) times higher than that of TNT. This highlights the significance of doping as an effective approach to achieve stable and high-performance BP-N.

**ACKNOWLEDGEMENTS**

This work is supported by the National Natural Science Foundation of China (NSFC) under Grant of U2030114, and CASHIPS Director's Fund (Grant No. YZJJ202207-CX). The calculations were partly performed in Center for Computational Science of CASHIPS, the ScGrid of Supercomputing Center and Computer Network Information Center of Chinese Academy of Sciences, and the Hefei Advanced Computing Center.

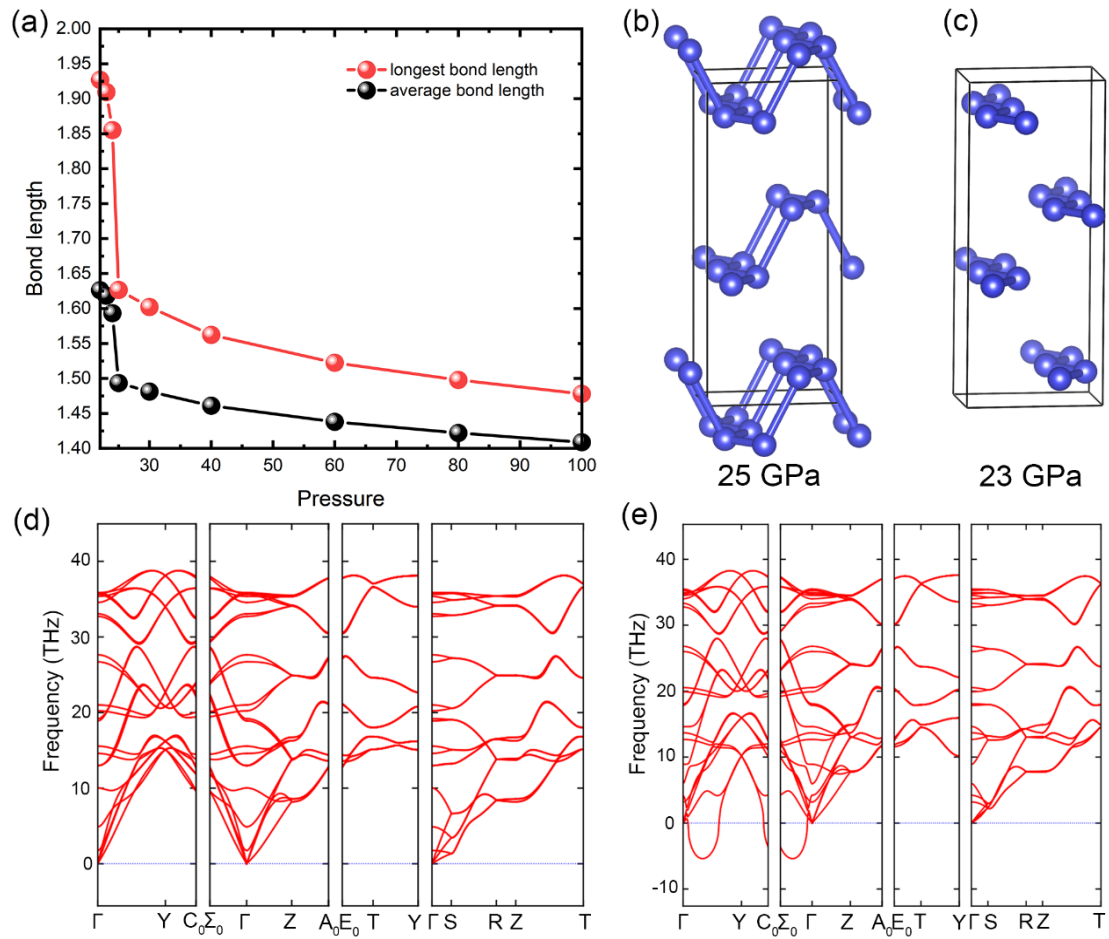

**FIG. 1.** (a) Longest (red circles) and Average (black circles) bond lengths of BP-N variations as the pressure. The relaxation proceeds from high pressure to low pressure, leading to an increase in bond lengths as pressure decreases. (b) BP-N' structure at 25 GPa. (c) BP-N' structure at 23 GPa. It could maintain stability at 25 GPa while decomposed into N-chains at 23 GPa due to the longest bond breaks. (d) The phonon spectrum of BP-N at 40 GPa. (e) The phonon spectrum of BP-N at 30 GPa.

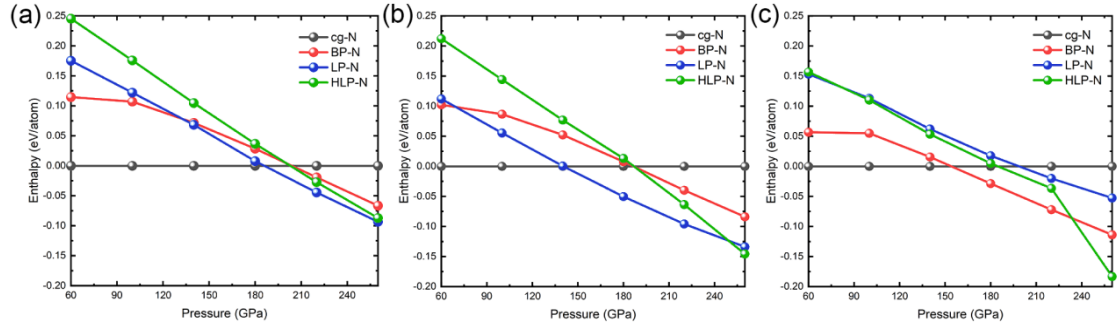

**FIG. 2.** Enthalpy variations at pressures of different P atom doping concentrations for different phases: cg-N, BP-N, LP-N, and HLP-N. (a) Configurations without P atom doping. (b) Configurations with 3.125 % doping concentration. (c) Configurations with 6.25 % doping concentration. The enthalpy values are referenced to the cg-N. For 3.25 % doping concentration, there are 1, 1, 4, 4 unique configurations for cg-N, BP-N, LP-N, and HLP-N, respectively, only stable configurations with minimal enthalpy were used. There is the same way for 6.25 % doping concentration.

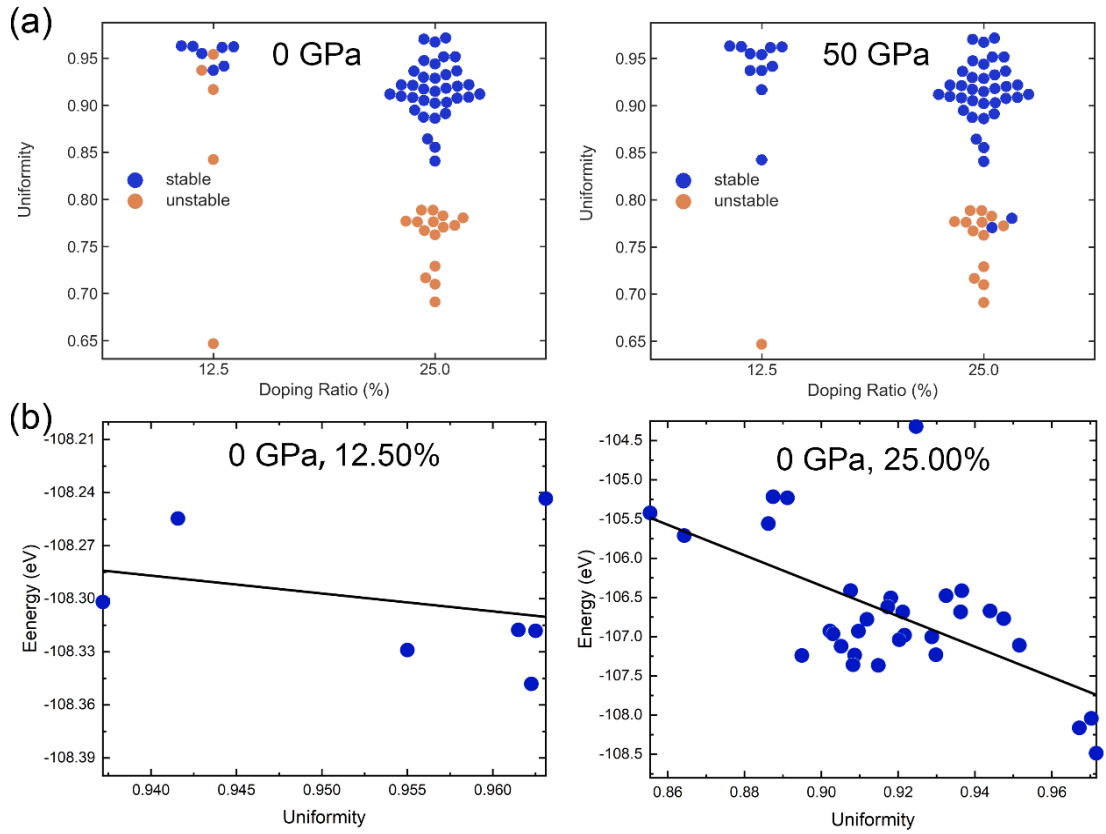

**FIG. 3.** Stability and energy changes observed across uniformity for 12.5% and 25% doping concentrations. (a) The relationship between stability and uniformity at 0 GPa and 50 GPa. The blue circles represent stable structures, while the orange circles depict unstable structures for structure relaxation. (b) The relationship between energy and uniformity at 0 GPa for 12.5% and 25% doping concentrations. All circles represent stable configurations, while the black lines represent guiding lines.

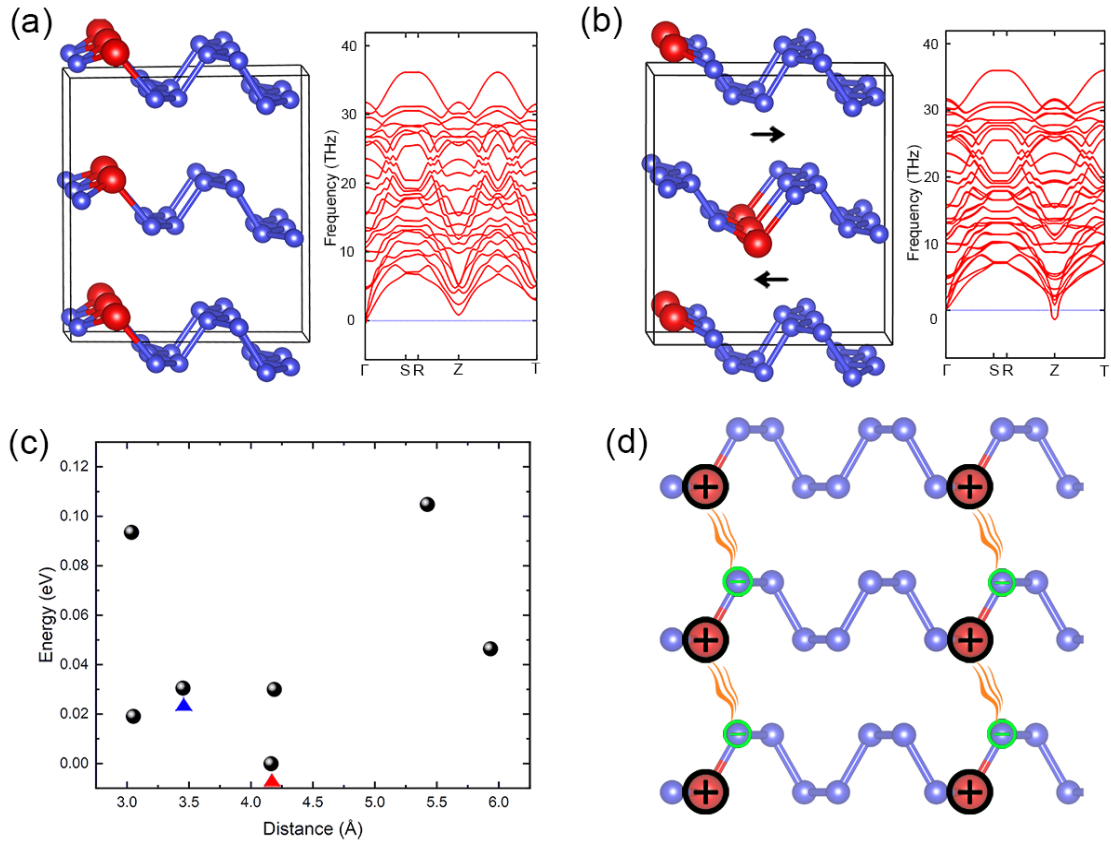

**FIG. 4.** Structures alongside their respective phonon dispersions for (a) the configuration possessing the lowest energy, and (b) a representation of one dynamically unstable configuration. The black arrows indicate the direction of vibration for the Z point's imaginary frequency. (c) Energy variations are depicted as a function of interlayer P-P distances. Each circle represents a unique configuration, wherein the red triangle denotes the configuration with the lowest energy, while the blue triangle represents one of the dynamically unstable configurations. (d) Schematic diagram of interlayer electrostatic interaction for configurations with the N-P dipoles. Atoms with charged labels (positive or negative) contribute to interlayer stability. The orange waves visually depict the electrostatic interaction between the P and N atoms, highlighting the key role it plays in stabilizing the system. For all structures, blue spheres represent N atoms, while red spheres represent P atoms.

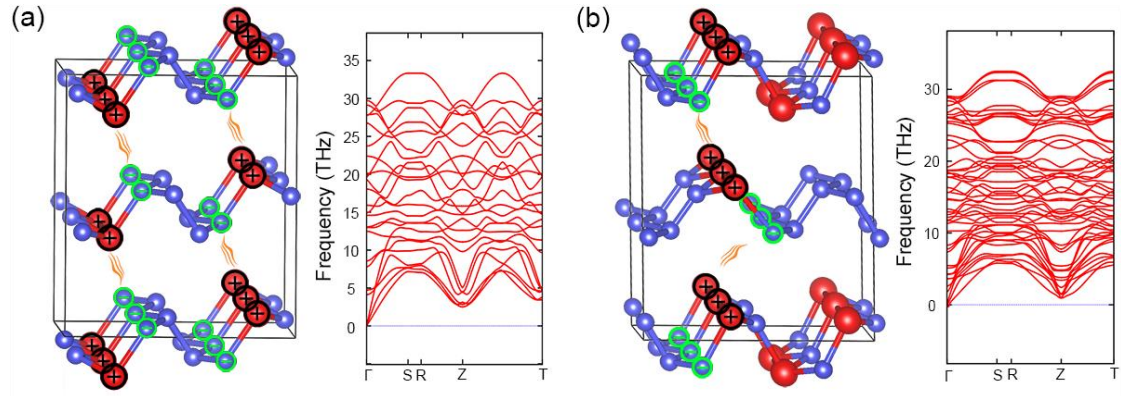

**FIG. 5.** Structures and corresponding phonon dispersions of 25% doping concentration. (a) Configuration with the four N-P dipoles, (b) Configuration with two N-P dipoles.. The orange waves visually depict the electrostatic interaction between the P and N atoms. Blue spheres represent N atoms, red spheres represent P atoms, and atoms with charged labels contribute to interlayer stability.

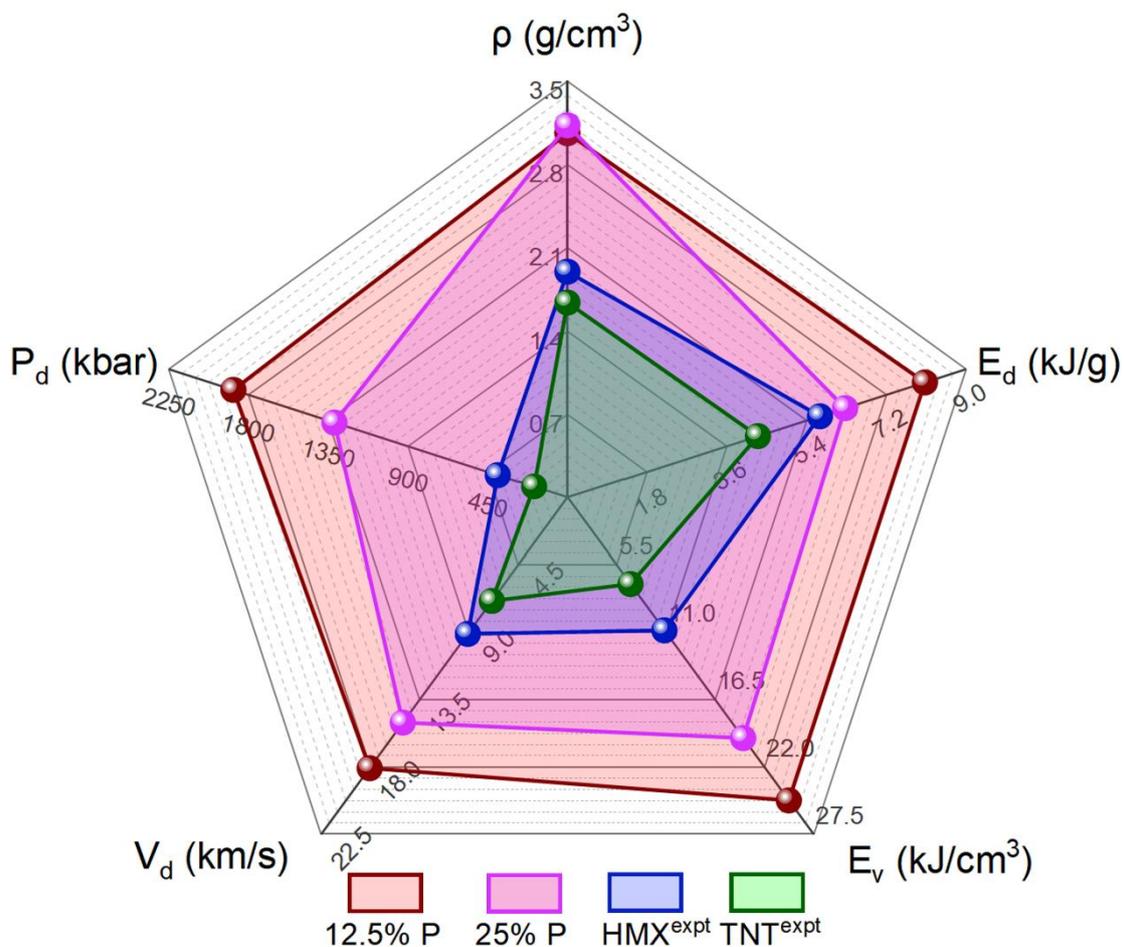

**FIG. 6.** Calculated density ($\rho$), energy density ($E_d$), volumetric energy density ($E_v$), detonation velocity ($V_d$), and detonation pressure ($P_d$) of 12.5% and 25% P atom doped BP-N. For comparison, TNT and HMX explosives are also listed. Herein, superscript expt represents the experimental data.